\documentclass{article}
\usepackage{amssymb}
\usepackage{amsmath}
\usepackage{graphicx}
\usepackage{graphics}
\usepackage{appendix}

\input{tcilatex}
\begin{document}

\title{Study of semileptonic decays of $D_{s}$ meson within R-parity
violating supersymmetric model.}
\author{\ $Farida\ Tahir\footnotemark ,~Azeem\ Mir\footnotemark $. \\
%EndAName
($\ast $\textit{Physics Department, COMSATS Institute of Information
Technology, Islamabad,}\\
$\dag $\textit{Physics Department, COMSATS Institute of Information
Technology, Lahore})}
\date{}
\maketitle

\begin{abstract}
\textit{We present the comaprative study of semileptonic and leptonic decays
of }$D_{s},D^{\pm }~$and $D^{0}$ meson $(D\rightarrow M\ l_{\alpha
}^{+}l_{\beta }^{-},D\rightarrow l_{\alpha }^{+}l_{\beta }^{-},D\rightarrow
l_{\alpha }^{+}v_{\alpha };\alpha ,\beta =e,\mu )\ $within\textit{\ the
framework of R-parity violating (}$\NEG{R}_{p}$\textit{) Minimal
Supersymmetric Standard Model (MSSM). The comparison shows that combination
and product couplings (}$\lambda _{\beta i\alpha }\lambda _{ijq}^{^{\prime
}\ast }$ or $\lambda _{\beta qk}^{^{\prime }}\lambda _{\alpha jk}^{^{\prime
}\ast })$ contribution to the branching fractions of the said processes
rocesses(under consideration) is consistent or comparable to the
experimental measurements in most of the cases. However, there exists some
cases, where these contributions are suppressed but still comparable to that
of standard model.
\end{abstract}

\section{Introduction}

\footnotetext[1]{%
neutrino\_79@hotmail.com}\footnotetext[2]{%
farida\_tahir@comsats.edu.pk, ftahir@ku.edu} Flavor changing neutral
current(FCNC) involving decays of charm mesons have played a leading role in
flavor physics as well as in search for new physics\cite{REN}. These
processes involve neutral meson oscillations, radiative decays and leptonic
and semileptonic(Lepton flavor and number conserving) decays. Lepton flavor
and number violating (LFV) processes are also important in this regard and
have been explored for signs of new physics. They are allowed in\ the SM
through higher order diagrams similar to FCNC but assuming oscillation in
the virtual neutrino loop. These diagrams are doubly cabibbo suppressed,
furthermore LFV are not allowed in SM. Therefore such processes provide the
golden charmed physics beyond the SM. Neutral meson oscillations like $D^{0}-%
\bar{D}^{0}$ mixing and coherent double-flavor oscillations is an exciting
new way to search for CP-violation in the charm sector\cite{NEUT}. Radiative
decays of charm mesons have been studied upto one loop order\cite{RAD}.
Processes involving $D$ meson have proven to be an excellent laboratory for
studying QCD since charm meson masses, O(2 GeV), are placed in the middle of
the region of non-perturbative hadronic physics\cite{PERT}.

New physics models, which are currently being explored in the charm sector
include Higgs models like 2HDM and Little Higgs model with T-parity\cite%
{Higgs}. R-parity violating MSSM Yukawa couplings have been used to study
anomaly in the branching fraction of leptonic decays of Ds meson, its
correlation with the branching fraction of $\tau $ LFV decays\cite{Rparit}.
The possibility of detecting signals of light sparticles using R parity
Yukawa couplings derived from the leptonic decays of $D_{S}$ meson has also
been explored\cite{Rparit}.Unparticle physics has also been explored in the
sector of charm mesons\cite{UNPART}.

Charm mesons($D,D^{\pm },D_{s})$ have been studied at facilities namely $%
E687 $, $E831\ ($Fermilab$),BES~III($Beijing Spectrometer III$)$,$CDF~$and
the $CLEO$ collaboration\cite{BES,ALEPH,FERMILAB}. Leptonic and semileptonic
decays of $D_{s}$ meson have also been studied in Belle and BaBar\cite{BELLE}%
.

The aim of this paper is to analyze FCNC processes involving leptonic and
semileptonic decays and compare them with current experimental limits. The
comparison is also made with the other decay processes of charm mesons
having the same quark subprocess. For example ($D_{s}^{\pm }\rightarrow
K^{\pm }l_{\alpha }^{+}l_{\beta }^{-},\ D^{0}\rightarrow l_{\alpha
}^{+}l_{\beta }^{-},\ D^{\pm }\rightarrow \pi ^{\pm }l_{\alpha }^{+}l_{\beta
}^{-}$) have the same subquark process ($c\rightarrow ul_{\alpha
}^{+}l_{\beta }^{-}$) while ($D^{0}\rightarrow \pi ^{-}l_{\alpha
}^{+}v_{\beta },\ D^{+}\rightarrow l_{\alpha }^{+}v_{\beta }$) have the same
subquark process ($c\rightarrow dl_{\alpha }^{+}v_{\beta }$). Spectator
quark model\cite{paula} is used to calculate branching fraction of above
mentioned processes.

FCNC proceed through box and penguin diagrams\cite{FERMILAB} within the
standard model(SM) and are highly suppressed\cite{BELLE,FCNC}. Since they
are allowed at tree level within R-parity violating MSSM\cite{Rparit} any
significant deviation from the SM prediction will hint at new physics. This
fact motivates us to study leptonic and semileptonic decays of charm mesons
within R-parity violating MSSM.

The Minimal Supersymmetric Standard Model\ ($MSSM$)\ \cite{SUSY}\ was
introduced by keeping in view the phenomenological implication of $SUSY$. It
contains the minimum number of particles and fields. It is also the minimal
extension of $SM$ having $N=1$ generators\cite{SUSY}.

MSSM allows processes that violate baryon and lepton number. It also allows
LFV processes. R-parity, a discrete symmetry is imposed to prevent baryon
number, lepton number and flavor violating processes. It is defined as $%
R_{p}=(-1)^{3B+L+2S}$\cite{H1}.\ R-parity conservation is phenomenologically
motivated and if relaxed carefully allows one to analyze rare and forbidden
decays while maintaining the stability of matter\cite{R. Barbie}. The
R-parity violating gauge invariant and renormalizable superpotential is\cite%
{H1}%
\begin{equation}
\ W_{\text{\textsl{$\NEG{R}$}}_{\text{\textbf{p}}}}=\frac{1}{2}\lambda
_{ijk}L_{i}L_{j}E_{k}^{c}+\lambda _{ijk}^{^{\prime }}L_{i}Q_{j}D_{k}^{c}+%
\frac{1}{2}\lambda _{ijk}^{^{\prime \prime
}}U_{i}^{^{C}}D_{j}^{^{C}}D_{k}^{^{C}}+\mu _{i}H_{u}L_{i},\ \ \ \ \ \ \ \ \ 
\end{equation}%
where $i,\,j,\,k$\ are generation indices, $L_{i}$\ and $Q_{i}$\ are the
lepton and quark left-handed $SU(2)_{L}$\ doublets and $E^{c}$, $D^{c}$\ are
the charge conjugates of the right-handed leptons and quark singlets,
respectively. $\lambda _{ijk}$, $\lambda _{ijk}^{\prime }$\ and\ $\lambda
_{ijk}^{\prime \prime }$ are Yukawa couplings. The term proportional to $%
\lambda _{ijk}$\ is antisymmetric in first two indices $[i,\,j]$\ and $%
\lambda _{ijk}^{\prime \prime }$ is antisymmetric in\ last$\ $two indices $%
[j,k]$, implying $9(\lambda _{ijk})+27(\lambda _{ijk}^{^{\prime }})+9\left(
\lambda _{ijk}^{^{\prime \prime }}\right) =45$\ independent coupling
constants among which 36 are related to the lepton flavor violation (9 from $%
LLE^{c}$\ and 27 from $LQD^{c}$). We can rotate the last term away without
affecting things of our interest.

\section{($D^{+},D_{s})\rightarrow l_{\protect\alpha }^{+}\protect\nu _{%
\protect\beta }$ In $\NEG{R}_{p}$ SUSY}

The effective Lagrangian for the decay of ($D^{+},D_{s})\rightarrow
l_{\alpha }^{+}+\nu _{\beta }$ in the quark mass basis is given as

\begin{equation}
L_{R_{P}\hspace{-0.38cm}/}^{eff}\left( c\longrightarrow q+l_{\alpha
}^{+}+\nu _{\beta }\right) =\frac{4G_{F}V_{cq}}{\sqrt{2}}\left[ 
\begin{array}{c}
A_{\alpha \beta }^{cq}\left( \bar{c}\gamma ^{\mu }P_{L}q\right) \left( 
\overline{l}_{\alpha }\gamma _{\mu }P_{L}\nu _{\beta }\right) \\ 
-B_{\alpha \beta }^{cq}\left( \bar{c}P_{R}q\right) \left( \overline{l}%
_{\alpha }P_{L}\nu _{\beta }\right)%
\end{array}%
\right] ,
\end{equation}%
where $\alpha ,\beta =e,\mu ~$and $q=d,s$. The dimensionless coupling
constants $A_{\alpha \beta }^{cq}$and $B_{\alpha \beta }^{cq}$ are given as,

\begin{eqnarray}
A_{\alpha \beta }^{cq} &=&\frac{\sqrt{2}}{4G_{F}V_{cq}}\underset{j,k=1}{%
\overset{3}{\sum }}\frac{_{1}}{2m_{\widetilde{d_{k}^{c}}}^{2}}V_{cj}\lambda
_{\beta qk}^{^{\prime }}\lambda _{\alpha jk}^{^{\prime }\ast }  \notag \\
B_{\alpha \beta }^{cq} &=&\frac{\sqrt{2}}{4G_{F}V_{cq}}\underset{i,j=1}{%
\overset{3}{\sum }}\frac{2}{m_{\widetilde{l_{i}^{c}}}^{2}}V_{cj}\lambda
_{\beta i\alpha }\lambda _{ijq}^{^{\prime }\ast }
\end{eqnarray}%
Thus the decay rate of the flavor conserving process $D^{+}\rightarrow
l_{\alpha }^{+}\nu _{\alpha }$ is given by

\begin{equation}
\Gamma \left( M^{-}\rightarrow l_{\alpha }\nu _{\alpha }\right) =\frac{1}{%
8\pi }G_{F}^{2}\mid V_{cq}\mid ^{2}f_{D}^{2}M_{D}^{3}\left( 1-\eta _{\alpha
}^{2}\right) ^{2}\mid (1+A_{\alpha \alpha }^{cq})\eta _{\alpha }-\left( 
\frac{M_{D}}{m_{c}+m_{d,s}}\right) B_{\alpha \alpha }^{cq}\mid ^{2}
\end{equation}%
where $\eta _{\alpha }=$ $\frac{m_{\alpha }}{M_{D}}$ is mass of charged
lepton $l$, $M_{D}$ is the mass of charm meson, where $f_{M}$ is
pseudoscalar meson decay constant. Here, following PCAC (partial
conservation of axial-vector current) relations have been used:

\begin{eqnarray}
&<&0\mid \overline{q}_{c}\gamma ^{\mu }\gamma _{5}q_{q}\mid
M(p)>=if_{M}p_{M}^{\mu }  \notag \\
&<&0\mid \overline{q}_{c}\gamma _{5}q_{q}\mid M(p)>=if_{M}\frac{M_{M}^{2}}{%
m_{q_{c}}+m_{q_{q}}}
\end{eqnarray}

The general decay rate including SM and R-parity violating contribution is
given by

\begin{equation}
\Gamma =\Gamma _{SM}(1+\alpha ).
\end{equation}

Where $\alpha $ is New Physics parameter (NP) given by

\begin{equation}
\alpha =\left\vert A_{\alpha \beta }^{cq}\right\vert ^{2}+\frac{1}{\eta
_{\alpha }^{2}}\left( \frac{M_{D}}{m_{c}+m_{d,s}}\right) ^{2}\left\vert
B_{\alpha \alpha }^{cq}\right\vert ^{2}  \label{1}
\end{equation}

The sneutrino Yukawa coupling products enhances SM contribution to the
branching fraction of leptonic decays ($D\rightarrow l_{\alpha
}^{+}v_{\alpha }$) many times\ ($\frac{1}{\eta _{\alpha }^{2}}$ in eq. (\ref%
{1})). So we do not include this factor to measure the NP.%
\begin{equation}
\alpha =\left\vert A_{\alpha \beta }^{cq}\right\vert ^{2}  \label{2}
\end{equation}

\section{$D\rightarrow (\protect\pi ,K)l_{\protect\alpha }^{+}\protect\nu _{%
\protect\beta }$ decay in $\NEG{R}_{p}$ SUSY}

The effective Lagrangian for the decay of $D\rightarrow (\pi ,K)l_{\alpha
}^{+}+\nu _{\beta }$ in the quark mass basis is given as

\begin{equation}
L_{R_{P}\hspace{-0.38cm}/}^{eff}\left( c\longrightarrow q+l_{\alpha
}^{+}+\nu _{\beta }\right) =-\frac{4G_{F}V_{cq}}{\sqrt{2}}\left[ 
\begin{array}{c}
A_{\alpha \beta }^{cq}\left( \bar{c}\gamma ^{\mu }P_{L}q\right) \left( 
\overline{l}_{\alpha }\gamma _{\mu }P_{L}\nu _{\beta }\right) \\ 
-B_{\alpha \beta }^{cq}\left( \bar{c}P_{R}q\right) \left( \overline{l}%
_{\alpha }P_{L}\nu _{\beta }\right)%
\end{array}%
\right] ,
\end{equation}%
where $\alpha ,\beta =e,\mu $ and $q=d,s$. The dimensionless coupling
constants $A_{\alpha \beta }^{cq}$and $B_{\alpha \beta }^{cq}$ are given as,

\begin{eqnarray}
A_{\alpha \beta }^{cq} &=&\frac{\sqrt{2}}{4G_{F}V_{cq}}\underset{j,k=1}{%
\overset{3}{\sum }}\frac{1}{2m_{\widetilde{d_{k}^{c}}}^{2}}V_{cj}\lambda
_{\beta qk}^{^{\prime }}\lambda _{\alpha jk}^{^{\prime }\ast }  \notag \\
B_{\alpha \beta }^{cq} &=&\frac{\sqrt{2}}{4G_{F}V_{cq}}\underset{i,j=1}{%
\overset{3}{\sum }}\frac{2}{m_{\widetilde{l_{i}^{c}}}^{2}}V_{cj}\lambda
_{\beta i\alpha }\lambda _{ijq}^{^{\prime }\ast }
\end{eqnarray}%
Thus the decay rate of $D\rightarrow Kl_{\alpha }^{+}\nu _{\beta }$ induced
by is given by

\begin{equation}
\Gamma \left[ c\rightarrow q\text{ }l_{\alpha }^{+}v_{\beta }\right] =\frac{%
m_{D}^{5}}{192\pi ^{3}}G_{F}^{2}\mid V_{cq}\mid ^{2}(\mid A_{\alpha \beta
}^{cq}\mid ^{2}+\frac{\mid B_{\alpha \beta }^{cq}\mid ^{2}}{4}).
\end{equation}

\section{$D^{0}\rightarrow l_{\protect\alpha }^{\pm }l_{\protect\beta }^{\mp
}$ In $\NEG{R}_{p}$ SUSY}

The effective Lagrangian for the decay of $D^{0}\rightarrow l_{\alpha }^{\pm
}l_{\beta }^{\mp }$ in the quark mass basis is given as

\begin{equation}
L_{\QTR{sl}{\NEG{R}\;}_{\text{\textbf{p}}}}^{eff}\left( c\longrightarrow
u+l_{\alpha }^{\pm }+l_{\beta }^{\mp }\right) =\frac{4G_{F}}{\sqrt{2}}\left[ 
\begin{array}{c}
A_{\alpha \beta }^{cu}\left( \overline{l_{\alpha }}\gamma ^{\mu
}P_{L}l_{\beta }\right) \left( \overline{u}\gamma _{\mu }P_{R}c\right)%
\end{array}%
\right] ,
\end{equation}%
where $\alpha ,\beta $ $=e,\mu .$The dimensionless coupling constants $%
A_{\alpha \beta }^{cu}$ is given by

\begin{equation}
A_{\alpha \beta }^{cu}=\frac{\sqrt{2}}{4G_{F}}\underset{i=1}{\overset{3}{%
\sum }}\frac{V_{ni}^{\dagger }V_{im}}{2m_{\widetilde{d_{i}^{c}}}^{2}}\lambda
_{\beta n1}^{^{\prime }}\lambda _{\alpha m2}^{^{\prime }\ast }  \label{a}
\end{equation}

The decay rate of the processes $M\rightarrow l_{\alpha }^{\pm }l_{\beta
}^{\mp }$ is given by

\begin{equation}
\Gamma \left[ M\left( cu\right) \rightarrow l_{\alpha }^{\pm }l_{\beta
}^{\mp }\right] =\frac{1}{8\pi }G_{F}^{2}f_{M}^{2}M_{M}^{3}\sqrt{1+\left(
\eta _{\alpha }^{2}+\eta _{\beta }^{2}\right) ^{2}-2\left( \eta _{\alpha
}^{2}+\eta _{\beta }^{2}\right) }\mid A_{\alpha \beta }^{cu}\mid ^{2}[(\eta
_{\alpha }^{2}+\eta _{\beta }^{2})-(\eta _{\alpha }^{2}-\eta _{\beta
}^{2})^{2}]
\end{equation}%
where $\eta _{\alpha }\equiv \frac{m_{\alpha }}{M_{M}}.$ $m_{\alpha }$ is
mass of lepton, $M_{M}$ is the mass of meson, $f_{M}$ is the pseudoscalar
meson decay constant which are extracted from the leptonic decay of each
pseudoscalar meson.

\section{$D_{s}\rightarrow Kl_{\protect\alpha }^{-}l_{\protect\beta }^{+}$
decay in $\NEG{R}_{p}$ SUSY}

In MSSM the relevant effective Lagrangian for the decay process $%
D_{s}\rightarrow Kl_{\alpha }^{-}l_{\beta }^{+}~$is given by\cite{FAR1}

\begin{equation}
L_{\QTR{sl}{\NEG{R}\;}_{\text{\textbf{p}}}}^{eff}\left( c\longrightarrow
u+l_{\alpha }^{-}+l_{\beta }^{+}\right) =\frac{4G_{F}}{\sqrt{2}}\left[ 
\begin{array}{c}
A_{\alpha \beta }^{cu}\left( \overline{l_{\alpha }}\gamma ^{\mu
}P_{L}l_{\beta }\right) \left( \overline{u}\gamma _{\mu }P_{R}c\right)%
\end{array}%
\right] .
\end{equation}%
Where $\alpha ,\beta =e,\mu .\ $The first term in eq. (2) comes from the up
squark exchange (where $c$ and $u$\ are up type quarks). The dimensionless
coupling constant $A_{\alpha \beta }^{cu}\ $is given by

\begin{equation}
A_{\alpha \beta }^{cu}=\frac{\sqrt{2}}{4G_{F}}\underset{m,n,i=1}{\overset{3}{%
\sum }}\frac{V_{nj}^{\dagger }V_{jm}}{2m_{\widetilde{d_{k}^{c}}}^{2}}\lambda
_{\beta n1}^{\prime }\lambda _{\alpha m2}^{\prime \ast },
\end{equation}%
The inclusive decay rate of the process is given by\cite{Ja}%
\begin{equation}
\Gamma \left[ c\rightarrow u\text{ }l_{\alpha }^{+}l_{\beta }^{-}\right] =%
\frac{m_{D^{+}}^{5}}{192\pi ^{3}}G_{F}^{2}\mid A_{\alpha \beta }^{cu}\mid
^{2}.
\end{equation}

\section{Results and Discussions}

We have plotted Figs.(4-13) using data\cite{PDG}. Table 1,2 and 3 summarizes
new bounds on the branching fraction of the given decay processes. We have
used the bounds on the Yukawa coupling products from $\cite{R. Barbie,BOUND}%
. $ In table 2 and 3, we have calculated our results on branching fraction
and Yukawa coupling bounds within 1$\sigma $ error.

Fig.(4) shows a comparison between $D^{\pm }\rightarrow \pi ^{\pm
}e^{+}e^{-} $ and $D^{0}\rightarrow e^{+}e^{-}.~$This comaprison shows that $%
\NEG{R}_{p}$ MSSM contribution to $D^{0}\rightarrow e^{+}e^{-}$ is
suppressed as compared to the current experimental limits. While a
comparison between $D^{\pm }\rightarrow \pi ^{\pm }e^{+}e^{-}$ and $%
D_{s}^{\pm }\rightarrow K^{\pm }e^{+}e^{-}\ $shows that $\NEG{R}_{p}$ MSSM
contribution to $D_{s}^{\pm }\rightarrow K^{\pm }e^{+}e^{-}$ is $10^{2}$
times smaller than the current experimental limits. So the experimental
limits on $D_{s}^{\pm }\rightarrow K^{\pm }e^{+}e^{-}$ is expected to get
lower.

Fig.(5) shows a comparison between $D^{\pm }\rightarrow \pi ^{\pm }\mu
^{+}\mu ^{-}$ and $D^{0}\rightarrow \mu ^{+}\mu ^{-}.~$This comaprison shows
that $\NEG{R}_{p}$ MSSM contribution to $D^{0}\rightarrow \mu ^{+}\mu ^{-}$
is $10^{2}$ times smaller than the current experimental limits. However,
this is significantly much better than the case of $D^{0}\rightarrow
e^{+}e^{-}$. This is because the branching fraction of the pure leptonic
decay depends directly to the square of lepton to meson mass ratio. A
comparison between $D^{\pm }\rightarrow \pi ^{\pm }\mu ^{+}\mu ^{-}$ and $%
D_{s}^{\pm }\rightarrow K^{\pm }\mu ^{+}\mu ^{-}\ $shows that $\NEG{R}_{p}$
MSSM contribution to $D_{s}^{\pm }\rightarrow K^{\pm }\mu ^{+}\mu ^{-}$ is
comparable to the experimental limits. So this is one decay process to be
explored at Fermilab and CLEO.

Fig.(6) shows a comparison between $D^{\pm }\rightarrow \pi ^{\pm }e^{+}\mu
^{-}$ and $D_{s}^{\pm }\rightarrow K^{\pm }e^{+}\mu ^{-}.~$This comparison
shows that $\NEG{R}_{p}$ MSSM contribution to $D_{s}^{\pm }\rightarrow
K^{\pm }e^{+}\mu ^{-}$ is similar to the current experimental limits.
Therefore, it is also a promising process to be explored at Fermilab and
CLEO.

Fig.(7) shows a comparison between $D^{0}\rightarrow \pi ^{-}e^{+}v_{e}$ and 
$D^{+}\rightarrow e^{+}v_{e}.~$This comparison shows that $\NEG{R}_{p}$ MSSM
contribution to $D^{0}\rightarrow \pi ^{-}e^{+}v_{e}$ is solely by squark
exchange Yukawa couplings ($\lambda _{\beta qk}^{^{\prime }}\lambda _{\alpha
jk}^{^{\prime }\ast }$) while $\NEG{R}_{p}$ MSSM contribution to $%
D^{+}\rightarrow e^{+}v_{e}$ is mostly by sneutrino exchange Yukawa
couplings ($\lambda _{\beta i\alpha }\lambda _{ijq}^{^{\prime }\ast }$). The
contribution to Br($D^{+}\rightarrow e^{+}v_{e}$)\ from squark exchange
Yukawa coupling products ($\lambda _{\beta qk}^{^{\prime }}\lambda _{\alpha
jk}^{^{\prime }\ast }$) is comparable with SM\ contribution ($\alpha \leq $%
15\%,see Fig.(12)) but negligible as compared to existing bounds. $%
D^{+}\rightarrow \pi ^{0}e^{+}v_{e}$ displays the same behavior similar to $%
D^{0}\rightarrow \pi ^{-}e^{+}v_{e}$ as shown in Fig. (8).

Fig.(9) displays a comparison between $D^{0}\rightarrow \pi ^{-}\mu
^{+}v_{\mu }$ and $D^{+}\rightarrow \mu ^{+}v_{\mu }.~$This comparison shows
that $\NEG{R}_{p}$ MSSM contribution to $D^{0}\rightarrow \pi ^{-}\mu
^{+}v_{\mu }$ is dominated by squark exchange The contribution to Br($%
D^{+}\rightarrow e^{+}v_{e}$) from squark Yukawa couplings ($\lambda _{\beta
qk}^{^{\prime }}\lambda _{\alpha jk}^{^{\prime }\ast }$) is comparable with
SM ($\alpha \leq $14\%,see Fig.(13)) while slepton exchange Yukawa couplings
($\lambda _{\beta i\alpha }\lambda _{ijq}^{^{\prime }\ast }$) exchange
Yukawa terms also contributes to $D^{+}\rightarrow \mu ^{+}v_{\mu }$.

Fig.(10) displays a comparison between $D^{0}\rightarrow K^{-}\mu ^{+}v_{\mu
}$ and $D_{s}^{+}\rightarrow \mu ^{+}v_{\mu }.~~$This comaprison shows that $%
\NEG{R}_{p}$ MSSM contribution to $D^{0}\rightarrow K^{-}\mu ^{+}v_{\mu }~$%
and $D_{s}^{+}\rightarrow \mu ^{+}v_{\mu }~$is consistent with available
experimental data. The contribution to Br($D^{+}\rightarrow \mu ^{+}v_{\mu }$%
) to from squark Yukawa coupling products to is comparable with SM ($\alpha
\leq $1.4\%,see Fig.(13))

Fig.(11) displays a comparison between $D^{0}\rightarrow K^{-}e^{+}v_{e}$
and $D_{s}^{+}\rightarrow e^{+}v_{e}.$This comaprison shows that $\NEG{R}_{p}
$ MSSM contribution to $D^{0}\rightarrow K^{-}e^{+}v_{e}$ is solely by
squark exchange Yukawa couplings ($\lambda _{\beta qk}^{^{\prime }}\lambda
_{\alpha jk}^{^{\prime }\ast }$) while $\NEG{R}_{p}$ MSSM contribution to $%
D_{s}^{+}\rightarrow e^{+}v_{e}$ is by slepton exchange Yukawa couplings ($%
\lambda _{\beta i\alpha }\lambda _{ijq}^{^{\prime }\ast }$). Table 3 also
shows that the contribution made by squark exchange Yukawa terms to the
branching fraction of ($D_{s}^{+}\rightarrow e^{+}v_{e})$ is suppressed but
is consistent with SM($\alpha \leq $10\%,see Fig.(13)).

Fig. (12-13) shows the variation of NP parameter $\alpha $(see eq.(\ref{1}%
)). The CKM factor $V_{cq}(q=d,s)$ is responsible for the higher and lower
value of $\alpha .\ $

\begin{equation*}
1.4\leq \alpha (\%)\leq 15.
\end{equation*}

The comparison in Table (1) shows that the branching fraction of some decay
processes like ($D^{0}\rightarrow e^{+}e^{-},D^{0}\rightarrow \mu ^{+}\mu
^{-},D_{s}^{\pm }\rightarrow K^{\pm }e^{+}e^{-})$ receives contribution from 
$\NEG{R}_{p}$ MSSM that is smaller than the current experimental limits.
While the decay processes ($D^{0}\rightarrow e^{+}\mu ^{-},D_{s}^{\pm
}\rightarrow K^{\pm }e^{+}\mu ^{-},D_{s}^{\pm }\rightarrow K^{\pm }\mu
^{+}\mu ^{-})$ receive sizeable contribution from $\NEG{R}_{p}$ MSSM that is
comparable to the current experimental limits. These processes are thus the
most important ones in the future searches of new physics in decays of charm
meson.

This comparison in Table (2) and Table (3) also shows that for the decay
processes ($D^{0}\rightarrow (\pi ,K)^{-}l_{\alpha }^{+}v_{\beta },\
D^{+}\rightarrow l_{\alpha }^{+}v_{\beta },D_{s}\rightarrow l_{\alpha
}^{+}v_{\beta }$), slepton exchange terms contribute the branching fraction
of leptonic decays only. The contribution made by squark exchange terms to
branching fraction of both leptonic and semileptonic decays is consistent
with experimental measurements in most of the cases. Table (2) also shows
that ($D^{0}\rightarrow \pi ^{-}e^{+}v_{e},D^{+}\rightarrow \pi
^{0}e^{+}v_{e},D^{0}\rightarrow \pi ^{-}\mu ^{+}v_{\mu },D^{+}\rightarrow
\mu ^{+}v_{\mu })$ are also very important ones in the future searches of
new physics in decays of charm meson.

Summarizing, we have analyzed decay processes ($D_{s}^{\pm }\rightarrow
K^{\pm }l_{\alpha }^{+}l_{\beta }^{-}(v_{\alpha }),\ D^{0}\rightarrow
l_{\alpha }^{+}l_{\beta }^{-},\ D^{\pm }\rightarrow \pi ^{\pm }l_{\alpha
}^{+}l_{\beta }^{-}(v_{\alpha })$) and compared their branching fractions
against a common parameter\ $\lambda _{\beta n1}^{\prime }\lambda _{\alpha
m2}^{\prime \ast }.$ The analysis shows that $D_{s}^{\pm }\rightarrow K^{\pm
}\mu ^{+}\mu ^{-}$ is the most favorable process within SM for study at
Fermilab and CLEO detector\cite{ALEPH,FERMILAB}. While $D_{s}^{\pm
}\rightarrow K^{\pm }e^{+}\mu ^{-}$ is the favorable process, not allowed by
SM to be searched for at these sites.

\subsubsection{\textbf{Acknowledgement}}

Azeem Mir and F.Tahir is indebted to the Higher Education Commission of
Pakistan for awarding the funds under grant- \# 20-577.

\begin{center}
\begin{tabular}{|c|c|c|c|c|}
\hline
Process & $%
\begin{array}{c}
\text{Subquark } \\ 
\text{Process}%
\end{array}%
$ & $%
\begin{array}{c}
\text{Branching} \\ 
\text{Fraction}%
\end{array}%
$ & $\NEG{R}_{p}\ $couplings & Branching Fraction \\ \hline
&  & (Experimental) &  & ($\NEG{R}_{p}\ $enhancement) \\ \hline
$D^{0}\rightarrow e^{+}e^{-}$ &  & $<1.2\times 10^{-6}$ & $\left\vert
\lambda _{131}^{\prime \ast }\lambda _{132}^{\prime }\right\vert $ & 
\TEXTsymbol{<}$2.8\times 10^{-12}($Weak$)$ \\ \cline{3-3}\cline{5-5}
$D_{s}^{\pm }\rightarrow K^{\pm }e^{+}e^{-}$ & $c\rightarrow u\ e^{+}e^{-}$
& $<1.6\times 10^{-3}$ &  & \TEXTsymbol{<}$4.6\times 10^{-5}($Weak$)$ \\ 
\cline{3-3}\cline{5-5}
$D^{\pm }\rightarrow \pi ^{\pm }e^{+}e^{-}$ &  & $<7.4\times 10^{-6}$ & $%
<1.98\times 10^{-3}$ & $<7.4\times 10^{-6}$Consistent \\ \hline
$D^{0}\rightarrow \mu ^{+}\mu ^{-}$ &  & $<1.3\times 10^{-6}$ & $\left\vert
\lambda _{231}^{\prime \ast }\lambda _{232}^{\prime }\right\vert $ & 
\TEXTsymbol{<}$6\times 10^{-8}($Weak$)$ \\ \cline{3-3}\cline{5-5}
$D_{s}^{\pm }\rightarrow K^{\pm }\mu ^{+}\mu ^{-}$ & $c\rightarrow u\ \mu
^{+}\mu ^{-}$ & $<3.6\times 10^{-5}$ &  & \TEXTsymbol{<}$2.4\times 10^{-5}($%
Comparable$)$ \\ \cline{3-3}\cline{5-5}
$D^{\pm }\rightarrow \pi ^{\pm }\mu ^{+}\mu ^{-}$ &  & $<3.9\times 10^{-6}$
& $<1.44\times 10^{-3}$ & $<3.9\times 10^{-6}$Consistent \\ \hline
$D^{0}\rightarrow e^{+}\mu ^{-}$ &  & $<8.1\times 10^{-7}$ & $\left\vert
\lambda _{231}^{\prime \ast }\lambda _{132}^{\prime }\right\vert $ & 
\TEXTsymbol{<}$4\times 10^{-7}($Comparable$)$ \\ \cline{3-3}\cline{5-5}
$D_{s}^{\pm }\rightarrow K^{\pm }e^{+}\mu ^{-}$ & $c\rightarrow u\ e^{+}\mu
^{-}$ & $<6.3\times 10^{-4}$ &  & \TEXTsymbol{<}$2.1\times 10^{-4}($%
Comparable$)$ \\ \cline{3-3}\cline{5-5}
$D^{\pm }\rightarrow \pi ^{\pm }e^{+}\mu ^{-}$ &  & $<3.4\times 10^{-5}$ & $%
<4.24\times 10^{-3}$ & $<3.4\times 10^{-5}$Consistent \\ \hline
\end{tabular}

Table 1: A table showing comparison between branching fraction of decay
processes of charmed mesons ($D_{s},\ D^{0},\ D^{\pm }$). Yukawa coulpings
are normalized to $1/(m_{\widetilde{d_{3}^{c}}}/100GeV)^{2}$

\newpage
\end{center}

\begin{tabular}{|l|l|l|l|l|}
\hline
Processes & $%
\begin{array}{c}
Subquark \\ 
\text{Process}%
\end{array}%
$ & $%
\begin{array}{c}
\text{Branching} \\ 
\text{Fraction}%
\end{array}%
$ & $\NEG{R}_{p}\ $couplings & Branching Fraction \\ \hline
&  & $%
\begin{array}{c}
(\text{Experimental}) \\ 
(\text{Standard Model)}%
\end{array}%
$ &  & ($\NEG{R}_{p}\ $contribution) \\ \hline
$D^{0}\rightarrow \pi ^{-}e^{+}v_{e}$ &  & $(2.83\pm 0.17)\times 10^{-3}$ & $%
\begin{array}{c}
\lambda _{113}^{^{\prime }}\lambda _{1j3}^{^{\prime }\ast } \\ 
=%
\end{array}%
$ & $(2.83\pm 0.17)\times 10^{-3}$(Consistent) \\ \cline{1-1}\cline{3-3}
$D^{+}\rightarrow e^{+}v_{e}$ & $(c\rightarrow u\ e^{+}v_{e})$ & $%
\begin{array}{c}
<2.4\times 10^{-5} \\ 
1.18\times 10^{-8}%
\end{array}%
$ & $%
\begin{array}{c}
\begin{array}{c}
-(6.2\pm 0.18)\times 10^{-2} \\ 
(6.2\pm 0.18)\times 10^{-2} \\ 
(\alpha \leq 15\%)%
\end{array}%
\end{array}%
$ & $%
\begin{array}{c}
(3.31\pm 0.14)\times 10^{-8} \\ 
(1.86\pm 0.03)\times 10^{-8}%
\end{array}%
\begin{array}{c}
(\text{consistent} \\ 
\text{with SM)}%
\end{array}%
$ \\ \cline{1-1}\cline{3-3}
$D^{+}\rightarrow \pi ^{0}e^{+}v_{e}$ &  & $(4.4\pm 0.7)\times 10^{-3}$ &  & 
$(4.4\pm 0.7)\times 10^{-3}$(Consistent) \\ \hline
$D^{0}\rightarrow \pi ^{-}\mu ^{+}v_{\mu }$ & $(c\rightarrow d\ \mu
^{+}v_{\mu })$ & $(2.37\pm 0.24)\times 10^{-3}$ & $%
\begin{array}{c}
\lambda _{213}^{^{\prime }}\lambda _{2j3}^{^{\prime }\ast } \\ 
=%
\end{array}%
$ & $(2.37\pm 0.24)\times 10^{-3}$(Consistent) \\ \cline{1-1}
$D^{+}\rightarrow \mu ^{+}v_{\mu }$ &  & $%
\begin{array}{c}
(4.4\pm 0.7)\times 10^{-4} \\ 
5\times 10^{-4}%
\end{array}%
$ & $%
\begin{array}{c}
-(5.67\pm 0.29)\times 10^{-2} \\ 
(5.67\pm 0.29)\times 10^{-2} \\ 
(\alpha \leq 14\%)%
\end{array}%
$ & $%
\begin{array}{c}
(3.31\pm 0.14)\times 10^{-4} \\ 
(7.6\pm 0.2)\times 10^{-4}%
\end{array}%
$(Consistent) \\ \cline{1-2}\hline\cline{2-3}
$D^{0}\rightarrow K^{-}e^{+}v_{e}$ & $(c\rightarrow s\ e^{+}v_{e})$ & $%
(3.58\pm 0.06)\%$ & $%
\begin{array}{c}
\lambda _{123}^{^{\prime }}\lambda _{1j3}^{^{\prime }\ast } \\ 
=%
\end{array}%
$ & $(3.58\pm 0.06)\%$(Consistent) \\ \cline{1-1}\cline{3-3}
$D_{s}^{+}\rightarrow e^{+}v_{e}$ &  & $%
\begin{array}{c}
<1.3\times 10^{-4} \\ 
1.5\times 10^{-7}%
\end{array}%
$ & $%
\begin{array}{c}
-(2.43\pm 0.21)\times 10^{-1} \\ 
(2.43\pm 0.21)\times 10^{-1} \\ 
(\alpha \leq 10\%)%
\end{array}%
$ & $%
\begin{array}{c}
(6.99\pm 0.06)\times 10^{-8} \\ 
(2.64\pm 0.01)\times 10^{-7}%
\end{array}%
\begin{array}{c}
(\text{consistent} \\ 
\text{with SM)}%
\end{array}%
$ \\ \cline{1-2}\hline\cline{2-3}
$D^{0}\rightarrow K^{-}\mu ^{+}v_{\mu }$ & $(c\rightarrow s\ \mu ^{+}v_{\mu
})$ & $(3.31\pm 0.13)\%$ & $%
\begin{array}{c}
\lambda _{223}^{^{\prime }}\lambda _{2j3}^{^{\prime }\ast } \\ 
=%
\end{array}%
$ & $(0.99\pm 0.81)\times 10^{-3}$(weak an order less) \\ 
\cline{1-1}\cline{3-3}
$D_{s}^{+}\rightarrow \mu ^{+}v_{\mu }$ &  & $%
\begin{array}{c}
(6.2\pm 0.6)\times 10^{-3} \\ 
6.5\times 10^{-3}%
\end{array}%
$ & $%
\begin{array}{c}
(-3.26\pm 1.74)\times 10^{-2} \\ 
(\alpha \leq 1.4\%)%
\end{array}%
$ & $(6.2\pm 0.6)\times 10^{-3}$(Consistent) \\ \hline
\end{tabular}%
\bigskip

Table 2: A table showing comparison between branching fraction of decay
processes of charmed mesons \newline
($D_{s},\ D^{0},\ D^{\pm }$).(*) indicates that R-parity contribution is
consistent with the experimental measurements. Squark Yukawa couplings
products are normalized as $1/(m_{\widetilde{d_{3}^{c}}}/100GeV)^{2}.$

\newpage

\begin{tabular}{|l|l|l|l|l|}
\hline
Processes & $%
\begin{array}{c}
Subquark \\ 
\text{Process}%
\end{array}%
$ & $%
\begin{array}{c}
\text{Branching} \\ 
\text{Fraction}%
\end{array}%
$ & $\NEG{R}_{p}\ $couplings & Branching Fraction \\ \hline
&  & $%
\begin{array}{c}
(\text{Experimental}) \\ 
(\text{Standard Model)}%
\end{array}%
$ &  & ($\NEG{R}_{p}\ $contribution) \\ \hline
$D^{0}\rightarrow \pi ^{-}e^{+}v_{e}$ &  & $(2.83\pm 0.17)\times 10^{-3}$ & $%
\begin{array}{c}
\left\vert \lambda _{3j1}^{\prime \ast }\lambda _{131}\right\vert \\ 
=%
\end{array}%
$ & $<1.51\times 10^{-6}($Weak$)$ \\ \cline{1-1}\cline{3-3}
$D^{+}\rightarrow e^{+}v_{e}$ & $(c\rightarrow u\ e^{+}v_{e})$ & $%
\begin{array}{c}
<2.4\times 10^{-5} \\ 
1.18\times 10^{-8}%
\end{array}%
$ & $<7.16\times 10^{-4}$ & $<2.4\times 10^{-5}$(Consistent) \\ 
\cline{1-1}\cline{3-3}
$D^{+}\rightarrow \pi ^{0}e^{+}v_{e}$ &  & $(4.4\pm 0.7)\times 10^{-3}$ &  & 
$<3.88\times 10^{-6}($Weak$)$ \\ \cline{1-2}\cline{4-5}
$D^{0}\rightarrow \pi ^{-}\mu ^{+}v_{\mu }$ & $(c\rightarrow u\ \mu
^{+}v_{\mu })$ & $(2.37\pm 0.24)\times 10^{-3}$ & $%
\begin{array}{c}
\left\vert \lambda _{3j1}^{\prime \ast }\lambda _{232}\right\vert \\ 
=%
\end{array}%
$ & $<(2.53\pm 0.44)\times 10^{-4}%
\begin{array}{c}
(\text{weak} \\ 
\text{an order less)}%
\end{array}%
$ \\ \cline{1-1}\cline{3-3}
$D^{+}\rightarrow \mu ^{+}v_{\mu }$ &  & $%
\begin{array}{c}
(4.4\pm 0.7)\times 10^{-4} \\ 
5\times 10^{-4}%
\end{array}%
$ & $<(9.24\pm 0.81)\times 10^{-3}$ & $<(4.4\pm 0.7)\times 10^{-4})$%
(Consistent) \\ \cline{1-2}\cline{1-1}\cline{3-5}\cline{3-3}
$D^{0}\rightarrow K^{-}e^{+}v_{e}$ & $(c\rightarrow s\ e^{+}v_{e})$ & $%
(3.58\pm 0.06)\%$ & $%
\begin{array}{c}
\left\vert \lambda _{3j2}^{\prime \ast }\lambda _{131}\right\vert \\ 
=%
\end{array}%
$ & $<9.79\times 10^{-6}($Weak$)$ \\ 
\cline{1-1}\cline{3-3}\cline{3-3}\cline{5-5}
$D_{s}^{+}\rightarrow e^{+}v_{e}$ &  & $%
\begin{array}{c}
<1.3\times 10^{-4} \\ 
1.5\times 10^{-7}%
\end{array}%
$ & $<1.82\times 10^{-3}$ & $<1.3\times 10^{-4}$(Consistent) \\ 
\cline{1-2}\cline{1-4}\cline{2-3}
$D^{0}\rightarrow K^{-}\mu ^{+}v_{\mu }$ & $(c\rightarrow s\ \mu ^{+}v_{\mu
})$ & $(3.31\pm 0.13)\%$ & $_{%
\begin{array}{c}
\left\vert \lambda _{3j2}^{\prime \ast }\lambda _{232}\right\vert \\ 
=%
\end{array}%
}$ & $<4.67\times 10^{-3}%
\begin{array}{c}
(\text{weak} \\ 
\text{an order less)}%
\end{array}%
$ \\ \cline{1-1}\cline{3-3}
$D_{s}^{+}\rightarrow \mu ^{+}v_{\mu }$ &  & $%
\begin{array}{c}
(6.2\pm 0.6)\times 10^{-3} \\ 
6.5\times 10^{-3}%
\end{array}%
$ & $<(3.78\pm 0.21)\times 10^{-2}$ & $<(6.2\pm 0.6)\times 10^{-3}$%
(Consistent) \\ \hline
\end{tabular}%
\newline

Table 3: A table showing comparison between branching fraction of decay
processes of charmed mesons \newline
($D_{s},\ D^{0},\ D^{\pm }$).(*) indicates that R-parity contribution is
consistent with the experimental measurements. Slepton Yukawa couplings
products are normalized as$~1/(m_{\widetilde{l_{3}^{c}}}/100GeV)^{2}.$

\end{document}